\documentclass[aps,prl,superscriptaddress,twocolumn,showpacs]{revtex4}
\usepackage{amsmath,epsfig,amssymb}
\usepackage{graphicx}

\newcommand{\ra}{\rangle}
\newcommand{\la}{\langle}

\begin{document}
\title{Capacitive coupling of atomic systems to mesoscopic conductors}

\author{Anders S.\ S\o rensen}
\affiliation{ITAMP, Harvard-Smithsonian Center for Astrophysics,
Cambridge, Ma 02138}
\affiliation{Physics Department, Harvard University,
Cambridge, Ma 02138} 

\author{Caspar H. van der Wal}
\affiliation{Physics Department, Harvard University,
Cambridge, Ma 02138} 

\author{Lilian I.\ Childress}
\affiliation{Physics Department, Harvard University,
Cambridge, Ma 02138} 

\author{Mikhail D.\ Lukin}
\affiliation{ITAMP, Harvard-Smithsonian Center for Astrophysics,
Cambridge, Ma 02138}
\affiliation{Physics Department, Harvard University,
Cambridge, Ma 02138} 

\begin{abstract}
  We describe a technique that enables a  strong, coherent coupling 
  between 
   isolated neutral atoms  and 
   mesoscopic conductors. The coupling is achieved by exciting atoms trapped
  above the surface of a superconducting transmission line into 
  Rydberg states with large electric dipole moments, that
  induce voltage fluctuations in the transmission line. Using a
  mechanism analogous to cavity quantum electrodynamics 
  an atomic state can be transferred  to a long-lived mode of the 
  fluctuating voltage, atoms separated by millimeters can be entangled, 
  or the quantum state of 
  a solid state device can be mapped onto atomic or
  photonic states.   
\end{abstract}


\maketitle

Stimulated by the emerging ideas 
from quantum information science,
physical mechanisms that can facilitate a strong, coherent coupling of 
controllable quantum systems are currently being actively explored
\cite{fortschritte}. 
Among the leading systems being investigated are
the atomic
physics implementations with trapped atoms and ions \cite{sackett,atoms}.
These are very attractive in view of
the long coherence times and the well developed techniques
for  detecting and manipulating the
internal states. Furthermore, the quantum states of atoms 
can be reversibly mapped into light, thereby providing the
essential interface for quantum communication
\cite{rempe,kimble,kuzmich,vanderwal}. 
At the same time, solid state systems are also being
intensively investigated \cite{devoret,dot,charge,flux}, 
and it is believed that the advanced level of micro-fabrication techniques 
could allow for the scaling of these systems 
to a large number of quantum bits. 

This Letter describes a technique to combine the principal
advantages of the  atomic and solid state  systems. This is achieved 
by capacitively  coupling a superconducting wire to neutral
atoms trapped above its surface.
If the
atoms are temporarily excited into the Rydberg states their 
large electric dipole
moments produce a strong interaction with the charge fluctuations in
the conductor.   
 This interaction can lead to a coupling between atomic    
states and a single, long lived mode of excitation in the conductor in
a manner similar to 
cavity quantum electrodynamics.  
Compared to the traditional approach of microwave cavity QED
\cite{haroche,haroche-collision,walther}, however,  
transmission line resonators \cite{highQ} feature a stronger 
confinement of the electromagnetic field, thereby significantly
enhancing the relevant coupling strength.


 The use
of the large dipole moments of the 
Rydberg levels to construct 
quantum gates has already been proposed \cite{jaksch-rydberg}. 
In free space, however,  the interaction is limited to nearby
atoms because the dipole-dipole interaction falls off
very rapidly with the interatomic
distance ($L$) as $1/L^3$. 
Here, the conductor acts as waveguide for the interaction between the atoms,
and the interaction strength can    
decrease as slow as $1/\sqrt{L}$.  Furthermore, the same approach allows  
 to coherently couple atomic and solid state quantum bits, since the resonator
mode can interact strongly with electronic spins in quantum
dots \cite{dot,lily} or  
with the charge degree of freedom of superconducting Cooper pair boxes
\cite{noise,charge,girvin}.  This
can be used to facilitate the reversible mapping of solid state 
qubits to  atoms, or even to optical photons with single atoms or
atomic ensembles acting as 
mediators \cite{kimble,rempe,kuzmich,vanderwal}. Finally, the present  
ideas can  likely be extended to other atomic systems such as 
trapped ions \cite{sackett}. By exploiting the dipole moment associated with the movement of an ion \cite{cirac-nature}  and using small wires to connect ions trapped in separate traps, this could   provide a novel
route to a scalable ion trap quantum computer.

\begin{figure}[tb]
  \centering
  \includegraphics[width=8cm]{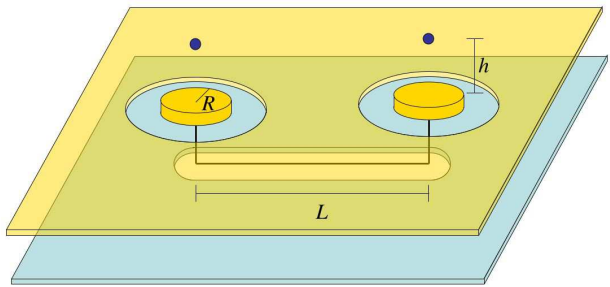}
  \caption{Considered experimental setup (not to scale).
  Two atoms are located at a
  height $h$ above two thin normal metal discs, which are connected by a
  thin superconducting wire of length $L$. In the shown coplanar waveguide
  geometry, the wire is screened from radiative losses by a
  superconducting plane (lower plane).  
  The discs and a normal metal layer (top plane) screen the wire from light 
  used for trapping and manipulating the atoms, such that no stray
  light is absorbed by the superconducting wire. }  
 \label{fig:setup}  
\end{figure}

 We consider the situation shown in 
Fig. \ref{fig:setup}. A single atom A 
(or a small atomic ensemble) is trapped at a distance 
$h$ above a conducting disc of radius $R$, 
which in turn is connected by a thin superconducting wire to a similar disc.  
When atom A is excited into a state with a large  dipole moment, it   
induces a charge $q_A$  on the disc below it and the opposite charge 
$-q_A$ is distributed on the wire and
the other disc. The second disk can be capacitively coupled to another
atom B, or 
alternatively,  to a solid-state quantum bit.  The charge induced by
atom A on the disc below atom B 
produces an electric field which interacts  with the dipole moment of
atom B, and produces an effective interaction between the two
atoms.
This interaction can be either electrostatic or 
electrodynamic in nature. In the electrostatic case, the dynamics
of the atomic dipole are much slower than the resonance frequency of
the wire so that the conductor adiabatically
follows the atomic states and produces a static
coupling between the atoms. 
In the electrodynamic case the resonance frequency of
the atomic dipole is tuned to resonance with one of the modes of the
wire producing a stronger and more flexible mechanism of
interaction.

If the dynamics of the system are much slower than the resonance frequency
of the wire ($\sim L/v$, where $v$ is the transmission velocity in the
wire) we can
neglect the inductance and the 
conductor is described by the Hamiltonian
$H_c=(q_A^2+q_B^2)/2C_d+(-q_A-q_B)^2/2C_w$, where $C_d$ ($C_w$) is the
capacitance of the discs (wire) and $q_A$ ($q_B)$ is the charge on the
disc below atom A (B). 
Assuming that the interaction between the wire and the atoms is
negligible, the interaction with atom $j$ ($j=$A or B) is given by 
$H_j=\vec{E}_j\cdot \vec{d}_j$, where $\vec{E}_j$ is the electric
field from the charge $q_j$ and $\vec{d}_j$ is the dipole operator of
atom $j$. Above the center of the disc the field is perpendicular to
the disc (along the $z$ axis) and is given by
 $E_{z,j}=q_j/(R^2+z^2)$ \cite{jackson} so that the interaction
Hamiltonian is 
\begin{equation}
  \label{eq:hab}
  H_j=\frac{q_j d_{z,j}}{R^2+h^2},
\end{equation}
where $d_{z,j}$ is the $z$ component of the dipole operator. Assuming
that the charge on the conductor  
follows the atomic dipoles adiabatically, $d H /d
q_j=0$ ($H=H_c+H_A+H_B$), we can eliminate the charges and  
obtain an effective dipole-dipole interaction between the two
atoms
\begin{equation}
  \label{eq:dipolfull}
  H_{{\rm int}}= \frac{d_{z,{\rm A}}d_{z,{\rm B}}}{(R^2+h^2)^2}
  \frac{C_d^2}{C_w+2C_d}.
\end{equation}
The capacitance of a disc is $C_d=2R/\pi$ \cite{jackson}, and to
estimate the 
obtainable coupling we take the capacitance of the 
wire to be given by the expression for a coaxial cable
$C_w=L/2\ln(b/a)$, where $a$ and $b$ are the inner and outer diameter
of the cable, and assume $\ln (b/a)\approx 1$.
In the limit $L\gg R$ with a fixed height $h$ the largest
coupling is achieved with 
$R=h$ and we find
\begin{equation}
  \label{eq:dipolsimple}
  H_{{\rm int}}=\frac{2}{\pi^2} \frac{d_{z,{\rm A}}d_{z,{\rm B}}}{h^2
  L}. 
\end{equation}
Compared with the interaction in free space $H_{{\rm
int}}\sim d_{z,{\rm A}}d_{z,{\rm B}}/L^3$, this method provides a much
stronger coupling  if  the atoms are close to the
conductor $h\ll L$. 

A stronger and more flexible coupling is possible using an
electrodynamic mechanism.   
The finite waveguide supports a set of quantized modes of frequency
$\omega_n \approx n \pi v/L$, where $n$ is any integer. If the
frequency of one  
of the modes in the conductor is on resonance with a transition
between two Rydberg levels $|r_1\rangle$ and $|r_2\rangle$,
the conductor can induce a transition between the   
Rydberg levels accompanied by a change
of excitation in the
conductor, similar to an atom emitting into a cavity.  To achieve such
a resonant interaction the length of the wire has to be similar to the
wavelength of the Rydberg transition which is typically in the range of
mm or cm.

To estimate the obtainable coupling we quantize the modes in the
conductor. If the wire has a charge per unit length $\lambda (x)$ and
current $i(x)$, where $x$ is the coordinate along the wire, the
Hamiltonian for the conductor is  
\begin{equation}
  \label{eq:hamwire}
  H_c=\frac{q_A^2+q_B^2}{2C_d}+\int {\left(
  \frac{\lambda(x)^2}{2c}+\frac{l}{2} i(x)^2\right)} dx, 
\end{equation}
where $c$ and $l$ are the capacitance and inductance per unit length. 
This Hamiltonian is diagonalized by expanding $\lambda$ and $i$ on
sines and cosines and
imposing the boundary condition that the potentials on the discs are
the same as at the ends of the wire.
Using the equations of motion $\partial\lambda/\partial t=-\partial
i/\partial x$ and $\partial i/\partial t=-1/v^2 \partial
\lambda/\partial x$, where 
$v=1/\sqrt{lc}$, we identify the canonical momentum $p_n$. The
Hamiltonian (\ref{eq:hamwire}) can then be written as a sum of harmonic
oscillator Hamiltonians $H_n=1/2 m_n \omega_n ^2 q_n^2+p_n^2/2m_n$,
where 
$q_n$ is the charge on the discs associated
with mode $n$, 
and the ``mass'' is $m_n
\approx C_w/2C_d^2\omega_n^2$ (for $L\gg R$).
Finally, we quantize the system by introducing the annihilation operators
$\hat{a}_n=q_n\sqrt{m_n\omega_n/2\hbar}+i p_n/ \sqrt{2m_n\hbar\omega_n}$ with
commutation relations $[\hat{a}_n,\hat{a}_n^\dagger]=1$. 

To describe the coupling between the atoms and the conductor we again
use (\ref{eq:hab}). If we write the dipole operator as $\hat{d}_z=
d_z ( |r_1\rangle\langle r_2| + |r_2\rangle\langle r_1|)$ the coupling
Hamiltonian has a form familiar from  cavity QED   
\begin{equation}
  \label{eq:cavity}
  H=g {\left(|r_1\rangle\langle r_2| \hat{a}^\dagger
  +|r_2\rangle\langle r_1| \hat{a} \right)}, 
\end{equation}
where we have performed the rotating wave approximation and omitted
all modes which are far from resonance with the transition. 
For $L\gg R$ the optimal coupling is again achieved by choosing
$R=h$, and we get 
\begin{equation}
g=d_z\sqrt{\frac{2\hbar \omega }{\pi^2 h^2L}}.
\label{eq:g}
\end{equation}
For a fixed transition frequency $\omega$ the coupling  falls off
slowly with the distance ($1/\sqrt{L}$) and therefore permits  
strong coupling of atoms with a large separation.  Comparing
with the expression from cavity QED, $g=d\sqrt{2\pi \hbar\omega/V}$,
the effective mode volume is $V=\pi^3h^2L$. The 
 waist of a cavity mode cannot be confined to less than a wavelength and 
the present approach allows for a stronger coupling if $h$ is 
smaller than the wavelength. 

The coupling in Eq.\ (\ref{eq:cavity}) is well studied 
and there are several proposals on how it can be exploited,
e.g., for generating entanglement or  quantum computation
\cite{pellizzari,domokos}. For these applications,  it  is
essential that the quantized modes of the wire are well isolated from
the environment. This is not the case if the wire is in free space
because it would strongly radiate, and it is therefore 
necessary to screen the wire. 
Experimentally the construction of such isolated
microwave resonators is being pursued
and $Q$-factors of $10^6$ have already been achieved \cite{highQ}.

To get a quantitative estimate
for the elctrodynamic case we consider the
transition from the p-state with principle quantum number N and
vanishing angular momentum along the $z$-axis to the $N-1$
s-state, and we ignore the 
quantum defects and the
fine and hyperfine structure in the atoms. For $N\gg 1$ the
matrix element for this transition is 
$d_z\approx eN^2 a_0/3\sqrt{3}$ \cite{bethe}, where  $e$ is the electron
charge and $a_0$ is the Bohr radius. The coupling constant is
$g\approx\hbar \omega \sqrt{2\alpha/(3\pi)^3 n} \sqrt{v_0/v} N^2 a_0/h$,
where $\alpha$ is the fine structure constant, $n$ is the number of
the mode which is 
resonant, and $v_0$ is the speed of light in vacuum.
If we take $n=1$, $v \approx v_0$,  $N=50$
corresponding to $\omega$ on the order of $(2\pi) 50$ GHz or $L\sim
n\cdot 3$ mm, and   
$h=10\ \mu$m we get $g\approx (2\pi)\hbar\  3$ MHz.  
For comparison, the decoherence rate associated with Rydberg
excitations can be in the range of kHz \cite{walther}, and the 
decoherence rate for an excitation in the conductor is $(2\pi)\
50$ kHz for  $Q \sim 10^6$. 

We next consider several practical aspects associated with the 
present technique.
 A major concern is the van der Waals forces on  
the atoms from the nearby conductor \cite{hinds}. Ideally one
would like to trap the Rydberg atoms to ensure that the motional
state remains decoupled from the evolution of the internal states. Such
trapping of Rydberg atoms has not yet been demonstrated, but even
without trapping the Rydberg levels it is still possible to
avoid a significant deterioration due to the motion of the atoms. If we
assume that the atoms are initially in their electronic ground state,
which can be 
trapped, e.g. by optical dipole traps or micro-fabricted magnetic traps \cite{chip1,chip2},
one can quickly excite the atoms into the Rydberg levels with a
resonant pulse, let them interact with the wire, and quickly deexcite
them after the interaction. A simple estimate of the force can be
found from the energy of a dipole above a plane $\Delta E=-\la
2\hat{d}_z^2 + \hat{d}_\rho^2\ra/16h^3$, where $\hat{d}_\rho$ is the
component of the dipole operator perpendicular to the $z$ axis.
For the numerical example considered above this gives
energy shifts $\Delta E \sim -(2\pi)\hbar\ 1$ MHz. Adding the resonant
contribution  we find
that the maximal force experienced by an atom is 
$F\leq(3\Delta E-g)/h$. If we assume that
the atoms are  initially cooled to the motional ground state of a trap
with a trapping  
frequency $\nu$, and that they are temporarily transferred to an
untrapped Rydberg state 
for a time $t$, the probability to be excited out of the ground state
is  $P\approx F^2t^2/2\hbar M \nu+\nu^2t^2/8$, where $M$ is the mass of
the atom. For a Rb atom with $\nu=(2\pi)\ 50$ kHz this gives $P\approx
10^{-3}$ for an 
interaction time  $t=\hbar \pi/g$.

In addition to the van der Waals forces, the Rydberg atom will also be influenced by fluctuating potentials from the surface. In current experiments with small superconducting quantum devices, the coherence times are limited by such charge  
fluctuations in the solid-state environment \cite{noise}. 
These effects are not yet fully understood and the experimental investigation 
is complicated by the lack of probes which do not share
the same environment. The Rydberg atoms could be used as a sensitive probe of these fluctuations that is controllably separated from the device itself.
If the atoms are excited into a state with a large angular momentum  where the Stark shift is linear, a single electron charge  added to an
island of $R=10\ \mu$m can shift the frequency  of an $N=50$ Rydberg
atom located 10 $\mu$m away  by 340 MHz. These effects are, however, much less important for the electrodynamic coupling since the non-degenerate s- and p-states experience a much weaker quadratic Stark shift. Noise in the solid states environment will cause voltage fluctuations of the wire of order $\langle V^2 \rangle \sim k_b T/C_w$, where $T$ is the temperature, and we estimate that these fluctuations plays a negligible role because of the large capacitance of the wire. In addition  the so called  patch potentials will influence the atoms \cite{patch}. As a specific example we estimate that the measured patch potentials in Ref.\ \cite{hinds}  would cause an energy shift of $\sim 7$ MHz at $h= 10\  \mu$m. This shift however can be compensated as long as it doesn't change in time, e.g., due to the movement of impurities on the surface. At low temperatures such  large-scale motions freeze out, thus reducing  the patch field fluctuations to an acceptable level for the electrodynamic experiments.

 

One of the main challenges for the experimental realization 
involves  combining strong optical
control fields (needed for trapping and manipulation of atoms) with 
the mesoscopic superconductors. If light is incident on
the superconductor the absorption of photons can break up a large
number of Cooper pairs. Even when heating effects are minimal,
this changes the inductance of the superconducting resonator
\cite{highQ},  
and cause decoherence of the oscillating mode. 
This problem can be avoided by  a design in which 
the discs at the ends of the transmission line are made of a normal
metal. The rest of the transmission line, the superconducting part, 
can be shielded by adding a metal layer as shown in
Fig.\ \ref{fig:setup}. Using here a normal metal-superconductor sandwich
combines low Ohmic losses for the bottom side with good cooling of the 
screening layer on the top side.
Note, that since the voltage on the disc
is determined by the total capacitance (which is mainly due to the long
wire) this metal layer does not affect the coupling strength
significantly. 
Because the current in the conductor makes a standing
wave with a node at the end, 
the finite conductivity of the discs does not affect the resonator
significantly. To estimate the obtainable $Q$ we replace the entire
conductor in Fig.\ \ref{fig:setup} with a wire of length $L=3$ mm
where the last 10 $\mu$m in both ends are made of gold. With a
conducting area of $\sim$1 $\mu{\rm m}^2$ the low resistance of
gold (for a thin layer at low temperature) makes a negligible
contribution ($Q \gtrsim 10^8$).
Another concern would be the resistance in the contacts
between the metals, but even with $\sim 0.1$ $\Omega$ contact resistance 
$Q \sim 10^7$. In practice radiative losses and loss in
the dielectric limit 
the ultimate value of the $Q$-factor.  Due to the pillars rising
above the plane the structure we propose in
Fig.\ \ref{fig:setup}, differs slightly from the structures
used for the experiments
\cite{highQ}. We estimate the radiative losses from these 
pillars to limit $Q$ to roughly $Q \lesssim (L/H)^4$, where $H$ is the
height of the pillars, and hence this modification has a negligible
contribution for $H\lesssim 30 \ \mu$m. In addition the radiative loss can
be suppressed by surrounding the setup with a Faraday cage, with the
top plane in Fig.\ \ref{fig:setup} as one of the walls. 

The above geometry allows to reduce heating effects from absorbing
stray light to an acceptable
level. With milliWatts of power in the beams focused on the atoms, 
stray light power at a level of nanoWatts can be
expected to be absorbed by the reflecting normal metal disc.
Even at an operating temperature of 100 mK , the 10 $\mu$m discs will 
easily have sufficient cooling power to the dielectric and the nearby
normal metal plane.   
Quasiparticle diffusion
into the superconducting wire can be minimized using a material with
fast quasiparticle relaxation and a
short coherence length (such as NbN), while increasing the dielectric
spacing
will further reduce the sensitivity of the inductance to the Cooper
pair density (desirable in contrast to \cite{highQ}). Although
sub-Kelvin regimes are desirable for low thermal occupation
of the oscillator modes, operation at temperatures as high as few Kelvin
might be feasible by using
atoms to cool the relevant modes \cite{haroche} or by involving
techniques analogous to
Ref.\cite{haroche-collision,sorensen}.

To summarize,  we have presented a feasible method to coherently couple atoms to a 
mesoscopic conductor with a coupling
strength significantly exceeding the decoherence rates.
It allows to combine solid-state micro-fabrication techniques with the  
excellent coherence properties and controllability of atomic
systems. Using the proposed technique atoms separated by millimeters can be 
entangled and one can imagine using this to  create 
scalable architectures for quantum computation 
where atoms are connected by wires. Also, the technique can be used to
couple solid state quantum bits to  
individual atoms or small atomic ensembles. The former 
can thereby be reversibly coupled to light. 
A different  application of the ideas presented here, could be the 
non-destructive measurement of the presence of a Rydberg atom by
 measuring the induced charge with a single electron transistor \cite{set}, or alternatively Rydberg atoms could be used to   probe the noise in solid state systems.

The full realization of the present proposal requires
a number of technological advances toward trapping and manipulation
of highly excited atoms in the vicinity of surfaces. 
Various elements of the proposed technique can, however,  be probed 
with existing technology. In particular, 
the basic mechanism of the capacitive coupling between atoms and  
transmission lines can be probed using a beam of highly excited atoms 
directed over the conductor
using the techniques demonstrated in Ref.\cite{hinds,at-sup}.

After submission of this work a related preprint appeared \cite{ionsolid} which describes an interface between a trapped ion and a solid state quantum computer using the static version of the  capacitive coupling. 

We are grateful to M.\ Brune, S.\ Haroche, J.-M.\ Raimond, G.\ Reithel, 
D.\ Vrinceanu, J.\ Zmuidzinas, and P.\ Zoller
for useful discussions.
This work was supported by the NSF 
through Career program and its
grant to ITAMP, ARO,  the Sloan and the  Packard Foundations, 
the Danish Natural Science Research
council, and the Hertz Foundation.

\end{document}